# Microstructure of water sediments on hydrophilic surfaces

**Tatyana Yakhno [1]\*, Anatoly Sanin [1] and Vladimir Yakhno [1,]**

[1] Federal Research Center Institute of Applied Physics of the Russian Academy of Sciences (IAP RAS), Nizhny Novgorod 603950, Russia
\* Correspondence: yakhta13@gmail.com (TY) Tel.: (011)-7-831-436-85-80

**Abstract:** In the modern world, the focus of natural science research thought has shifted mainly to the molecular level, including the study of water. Water is considered as a mixture of interacting $H_2O$ molecules and their clusters based on the data of molecular dynamics, neutron and X-ray scattering. In this case, a huge layer of information is lost related to the composition of real water, its microstructure and behavior. In this work, we studied the structure and dynamics of water during its interaction with glass surfaces during drying using optical microscopy. We observed a panoramic picture of a cascade of interrelated phase transitions of distilled water and its microdispersed phase - sodium chloride microcrystals surrounded by a thick layer of hydrate liquid crystal water. Incubation of a glass slide in water for 2 days was accompanied by adhesion of the microdisperse phase to the glass surface. During the free evaporation of distilled water from a Petri dish, erosion of hydration shells, dissolution of NaCl microcrystals, and further recrystallization with the formation of large monolithic crystals and traces of non-drying water were observed. However, there are still questions that do not yet have clear answers. The authors hope to continue research in collaboration with interested colleagues.

## 1. Introduction

The interaction of water with surfaces is the most important problem of physical chemistry underlying the development of animate and inanimate nature [1]. Despite the close attention of scientists to this topic and a number of undoubted discoveries in this area over the past decades, the scientific community is far from unanimous in interpreting the effects observed in the zone of water-surface contacts [2–6]. Therefore, the study of the microstructure of aqueous sediments on the surface in contact with water after the evaporation of free water, as well as the evolution of these sediments during drying in air, can clarify their origin and composition [7]. In the framework of this work, we studied the microstructure of the precipitate formed on a glass slide after its incubation in a glass of water (distilled or tap water) at room temperature for two days and subsequent drying of this precipitate in air. In another experiment, we expected the evaporation of water from Petri dishes poured into them with a layer 5 mm thick and placed in a turned off thermostat to protect them from room dust. After the formation of sediment at the bottom, we also observed its changes within 2 weeks.

Previously, the opinion was expressed [8] that "the adsorption of dissolved substances at the interfaces is due to direct interactions of the dissolved substance with the surface and solvation forces. These solvation forces are closely related to the structure of water in the interfacial region and are therefore determined by complex interactions in multidimensional space." In the conclusion of the article [8], the authors stated: "attempts to analytically describe the solvation forces at the interfaces are difficult and cannot be achieved in a predictive way." According to modern concepts [2], upon contact with hydrophilic surfaces, water forms a liquid-crystal near-wall layer up to 0.5 mm wide, as a result of which all dissolved substances and insoluble impurities are displaced from this "exclusion zone" (EZ) [2]. At the same time, due to the special supramolecular structure of near-wall water, EZ acquires a negative charge at the boundary with free water. Thus, in one glass under room conditions, two phases of water coexist - free and liquid crystal. It is shown that the distinguishing feature of liquid crystal water is its high electrical sensitivity [9] and higher refractive index [10] than that of ordinary water. The authors of [11] used a new phase-sensitive method to obtain an unprecedentedly detailed structure of the water/quartz interface. In their opinion, interfacial water molecules form a network of hydrogen bonds with mixed regions of ordered and disordered structures. Those located in ordered (disordered) regions, giving an ice-like (liquid-like) peak in the spectra, are closely associated with surface regions that have a higher (lower) pK value for deprotonation. According to cryogenic scanning electron microscopy, EZ water has a heterogeneous structure similar to a sponge, where high density water forms walls, between which there is low density water [12]. Hydrophilic microimpurities, getting into the water, also acquire hydrate shells (EZ). It is assumed that the interaction of such microparticles with each other and with near-wall water is carried out according to the "Like likes like" mechanism [13], i.e., through ions of the opposite sign (+) accumulating near EZ, which carries a negative charge [2].

Previously, we showed that real water in the real world always contains microimpurities, which are its integral part [14-16]. In this work, we intend to follow the fate of microimpurities during the formation of a deposit on the glass surface.

## 2. Materials and Methods

The work used distilled water GOST 6709-72 (pH 6,6; specific conductivity 5 μS/cm), distilled water "Water for injections" (manufacturer: Deko Company (Russia), LS-000512), as well as tap water (pH 7,2; specific conductivity 550 μS/cm).

New ApexLab microscope slides (25.4 x 76.2 mm, 1mm - 1.2 mm thick) and new Petry dishes (Russia) were washed in tap water and rinsed thoroughly in distilled water to re-move possible mechanical impurities. After drying the glasses in a vertical position on the table, they were used for experiments: they were immersed in a glass of water and left there for two days at room conditions: T = 23 - 25°C and H = 70 - 72%. After that, the slides were removed from the beakers and placed in a horizontal position on filter paper to evaporate free water. In the second series of experiments, water was poured into washed Petri dishes in a layer of 5 mm thick and left in a non-operating thermostat until free water completely evaporated (three days). The studies of dry sediments were performed under a Levenhuk microscope with a video camera coupled to a computer using the ToupView program without the use of a cover glass.

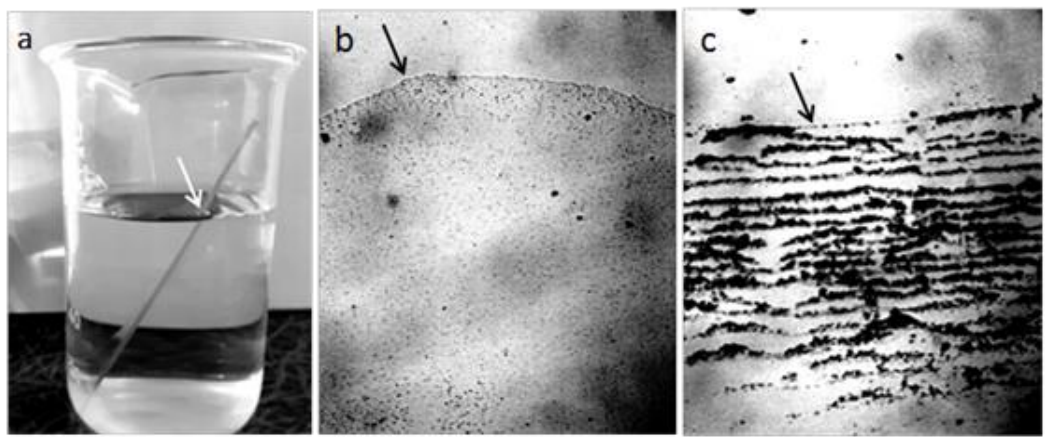

**Figure 1.** Scheme of the experiment (a) and the structure of deposits on glasses after two days of their incubation in distilled (b) and tap (c) water (microphoto). Frame height b, c - 3.0 mm.

As can be seen in the photographs, the deposits on the glass slide (Figure 1b) after its contact with distilled water are an even layer of densely packed rounded structures with a dark dot in the center, which we previously called the dispersed phase of water (DP) [7,14, 17]. We believe that the DP unit is formed by a hydrophilic microparticle (NaCl micro-crystal) covered with a thick layer of hydrate liquid crystal water (EZ). The upper part of the glass incubated in tap water (Figure 1c) is covered with bands of deposits of coarse microimpurities, reflecting the decreasing of the level of evaporating water. The deposition of DP aggregates in the middle part of the glasses can be multilayered (Figure 2). In sediments of distilled water after drying them in air, there were different-scale fragments of network structures, the optical density of which was higher than the back-ground (DP of water), and the cell sizes ranged from tens to hundreds of micrometers (Figure 3). On the one hand, cryogenic scanning electron microscopy suggests that EZ consists of high density water with a cell-like wall structure [12]. The paper reports that water has a heterogeneous structure consisting of water of high and low density. Using cryogenic scanning electron microscopy, the authors demonstrated the presence of a cel-lular structure where high density water exists as walls. Between them is ordinary water with a low density. From the measurement of electric charge separation, it was under-stood that the source of high density water is EZ.

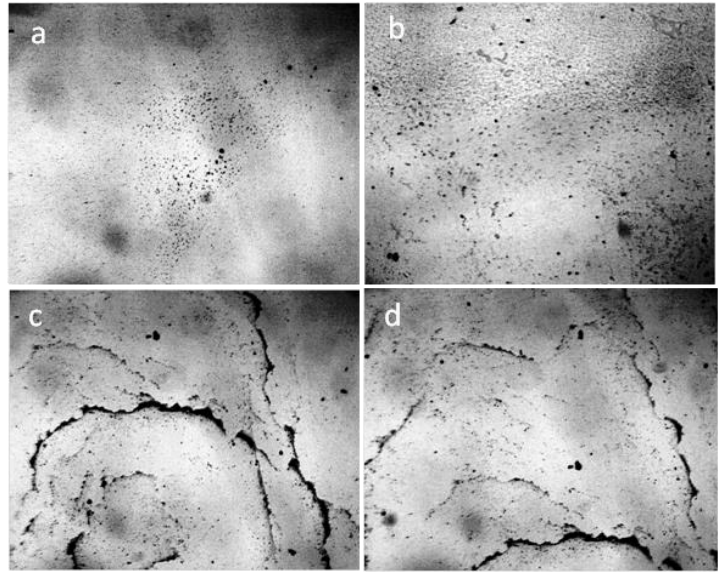

**Figure 2.** Microphoto. Middle part of incubated slides. a, b – deposition of DP aggregates in dis-tilled water; c,d are layers of DP aggregates in tap water. The width of each frame is 3 mm.

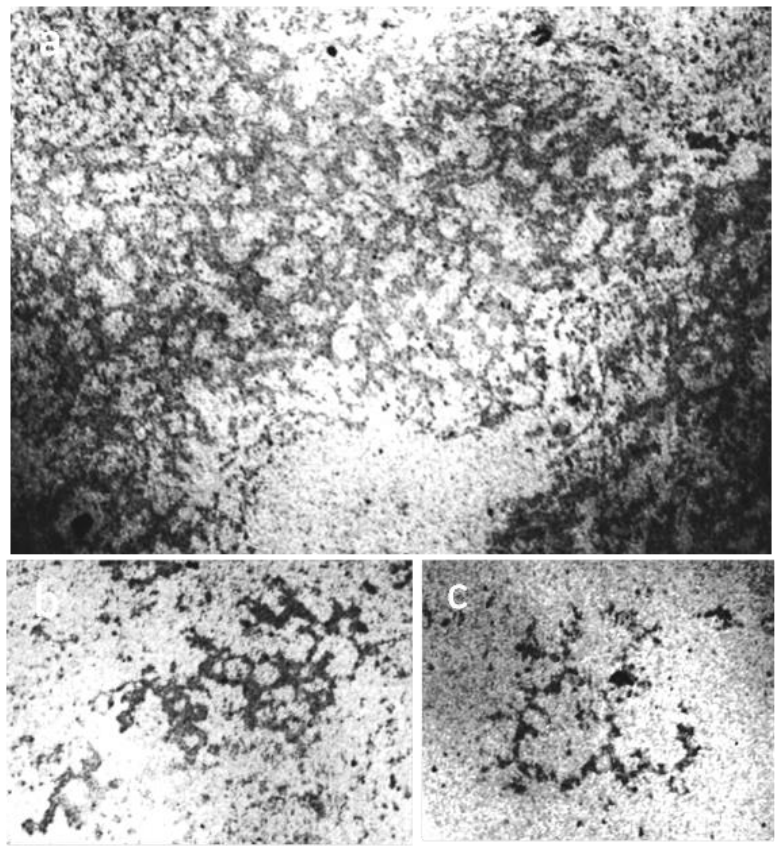

**Figure 3.** Microphoto. Net structures observed in dried sediments of distilled water after incubation of a glass slide in this water for 2 days: a - frame width - 3.0 mm; b - frame width - 1.0 mm; c - frame width - 0.5 mm.

The proposed model can adequately explain the anomalous properties, especially the electrical properties of water. As the size of the unit cell decreases, the "degree of struc-turedness" increases, which leads to an increase in the charging capacity of structured water due to an increase in the total interface area. At the same time, due to the separation of electrons, the redox potential and antioxidant property increase [12]. The size of the cells presented in this work was one to two orders of magnitude smaller than those observed by us in a light microscope. In the sediment of tap water at the bottom of the Petri dish, there were also mesh structures made of optically dense material (Figure 4). The cells limited the DP aggregates with growing inside them, according to our previous studies [7,14,15], sodium chloride crystals (Figure 5).

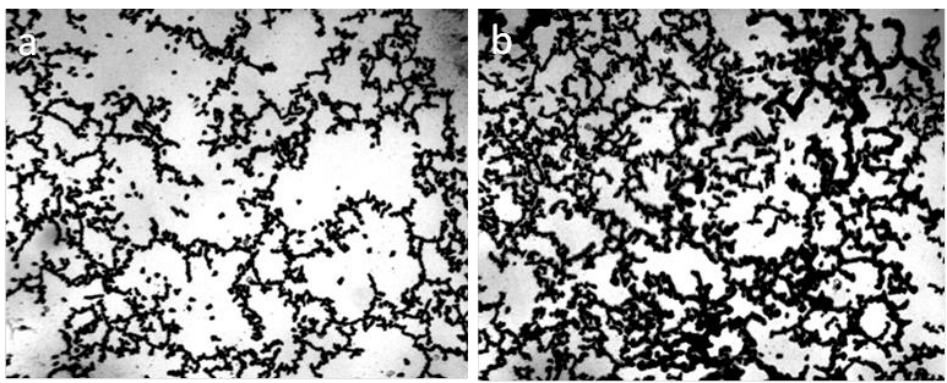

**Figure 4.** Microphoto. Fragments of tap water sediment in a Petri dish. a, b - different fields of view. Light areas bounded by dark boundaries are DP aggregates with NaCl crystals growing inside them (a). The width of each frame is 3.0 mm.

Arguing about the possible nature of "network structures", we would not exclude the trace of foam - a mass of air microbubbles distributed over the glass surface (Figure 5). It is clearly seen that the pattern on the glass strongly resembles the microscopic picture in Figure 3.

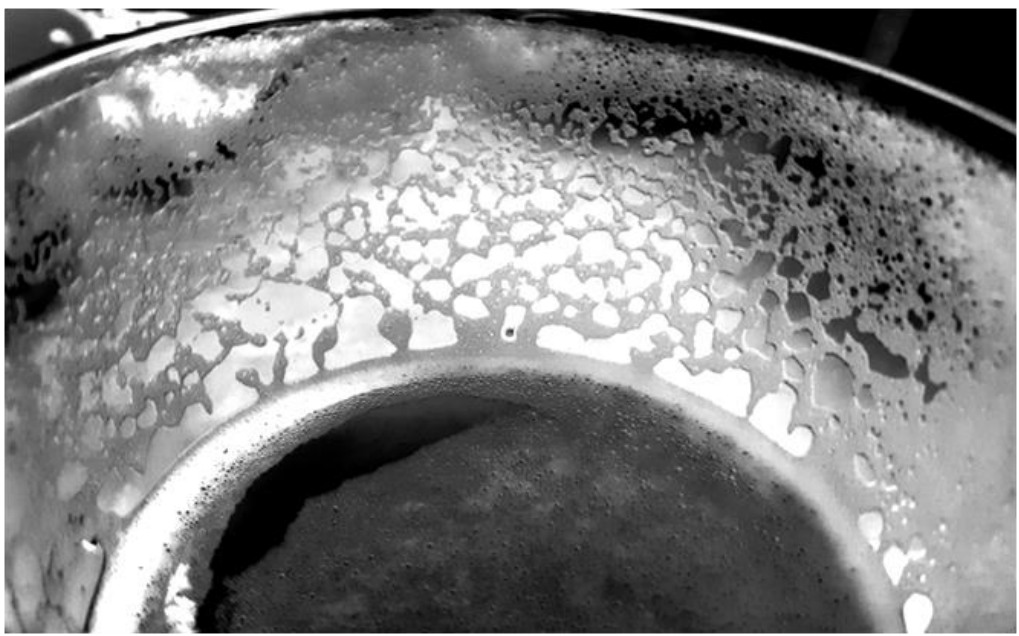

**Figure 5.** A pattern of foam residue on the wall of a beer glass.

Following this ("foamy") version, we must recall that, according to the literature [19], EZ is formed not only at hydrophilic surfaces, but also at the boundary with air. In solution, air bubbles are macroscopic hydrophobic surfaces. It is known that large hydrophobic regions form a thin layer of low density liquid immediately at the surface. In particular, the authors of [20] observed an increase in the areas of hexagonal packing as the level of solid particles in the droplets, and hence the elasticity, decreases. The layer can be associated with insta-bility and rupture of a thin liquid film between the bubbles [21]. Numerous studies have established that NaCl ions prevent bubble coalescence [22–27].

The evaporation of distilled water from the Petri dish was accompanied by a step-wise advance of the drying front (Figure 6).

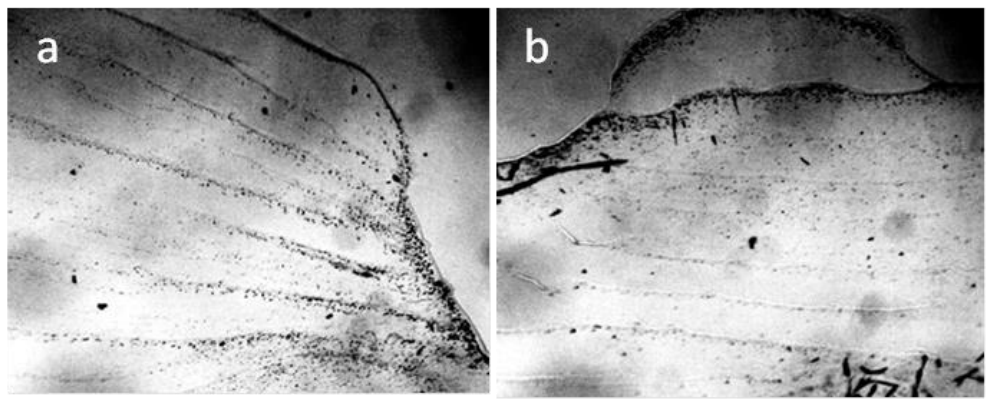

**Figure 6.** Microphoto. Dried precipitate of distilled water. Traces of stepwise advancement of the water drying front along the bottom of the Petri dish. a, b - different fields of view. The width of each frame is 3.0 mm.

Why does a layer of liquid under drying on a flat surface move not smoothly, but in the form of steps? The explanation for this was given by the Soviet physicist Ya. E. Geguzin (1973) [28] using the example of a drying drop of salt solution. Water from the entire sur-face of the drop evaporates evenly. As moisture evaporates, the concentration of dissolved salt will increase, and crystals will begin to fall out first where the excess salt concentration is greatest. This will be in the thinnest part of the drop - along its perimeter. The droplet liquid, wetting the falling out crystals, as if sticks to them. Therefore, a drop, losing liquid, must change its shape, become flatter - after all, its volume decreases, while the perimeter remains unchanged. What follows is explained by the competition between the increasing force of the surface and the force of adhesion of the liquid to the precipitated crystals. At a certain moment, the surface force overcomes the adhesion, and the water front jumps to a new step. The mechanism of deposit formation at the three-phase boundary of drying drops based on the Marangoni flow is also described in [29]. Thus, Figure 6 clearly shows that real distilled water is not a "pure liquid" but a microdispersion.

Let us consider the dynamics of phase transitions of distilled water components in accordance with the obtained data. This object is a system which consists of distilled wa-ter, freed from most salts, and a microdispersed phase consisting of NaCl microcrystals surrounded by a thick layer of hydrated liquid crystal water (EZ) [14,15]. Hydrated water has no solvent power, so the system is stable in the absence of evaporation. Evaporation is accompanied by an increase in the concentration of salt dissolved in the liquid phase, and, consequently, the osmotic pressure. Upon reaching a certain threshold of osmotic pressure, the hydration shells of DP begin to "melt" [15]. The liberated salt microcrystals get the op-portunity for further growth and "Ostwald maturation". On Fig. 7 shows a fragment of the DP sediment of distilled water at the bottom of the Petri dish at different depths of focus of the objective. It can be seen that microcrystals of sodium chloride protrude from the hydra-tion shells which are under destruction. In the immediate vicinity of them, large NaCl crystals are visible. The same large crystals can also be observed at the bottom of the Petri dish far from the sediments of DP water (Fig. 8). There is no doubt that these crystals were absent at the be-ginning of our experiment. Following the logic of our observations, the crystals were formed during the experiment due to the evaporation of free water. With increasing NaCl concentrations, the dissolved Na+ and Cl- ions tend to be aggregated in solutions. The for-mation of solute aggregate lowers the barrier height of nucleation and affects the nucleation mechanism of NaCl crystal in water [30]. The same large crystals can also be observed at the bottom of the Petri dish far from the sediments of DP water (Fig. 8). There is no doubt that these crystals were absent at the be-ginning of our experiment. Following the logic of our observations, the crystals were formed during the experiment due to the evaporation of free water. With increasing NaCl concentrations, the dissolved Na+ and Cl- ions tend to be aggregated in solutions. The for-mation of solute aggregate lowers the barrier height of nucleation and affects the nuclea-tion mechanism of NaCl crystal in water [30].

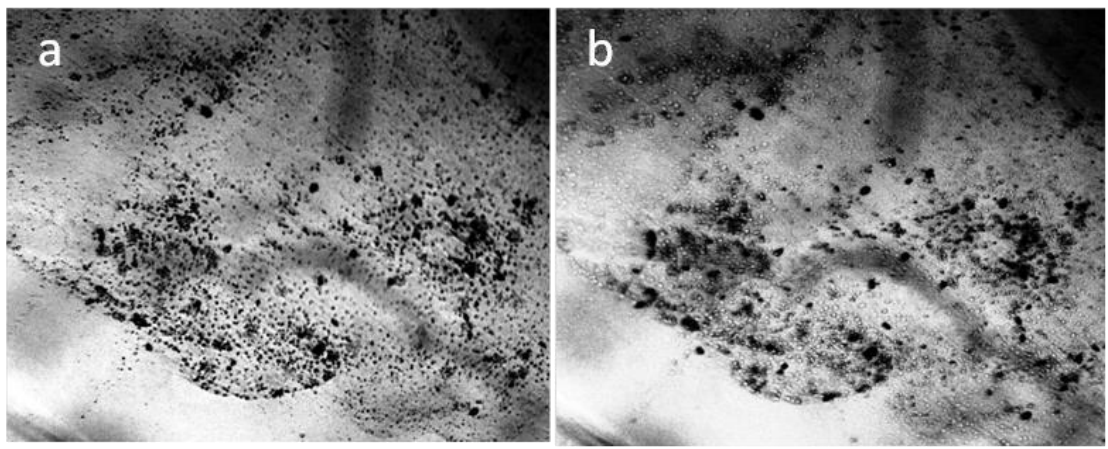

**Figure 7.** Fragment of the distilled water DF sediment at the bottom of the Petri dish: a – in normal viewing mode; b - with a decrease in the lens depth of field. Salt microcrystals (white dots) are visible, the height of which exceeds the average height of the preparation due to the partial dis-solution of hydration shells [15]. Microphoto. The width of each frame is 3 mm.

To illustrate the complex processes of phase transitions in distilled water drying on a glass surface, we consider it appropriate to present a diagram from our previous publication [15]. Our new results are consistent with previous ones.

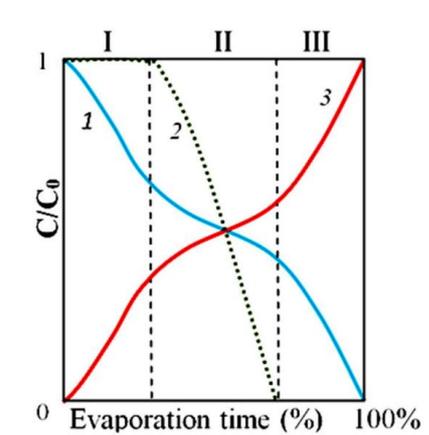

**Figure 8.** Schematic diagram of the dynamics of phase transitions in water containing Liquid Crystal Spheres with NaCl microcrystals as a "seed" during its evaporation from a solid substrate in the "Relative concentration–Evaporation time (%)" coordinates. Stage I—evaporation of free water and increase in osmotic pressure; Stage II—phase transition of liquid-crystal water into free water and reducing relative rate of evaporation; Stage III—growth of NaCl crystals. 1—relative concentration of free water; 2—relative concentration of liquid-crystal water; 3—relative salt concentration [15].

One more interesting fact attracts attention - the presence of non-drying water in the prep-arations even 2 weeks after the evaporation of distilled water (Fig. 9). Solutions of salts in water can be considered as solutions of hydrated ion complexes. It is shown that "the con-centration dependence of the verified solvent activity coefficient has the form of a discon-tinuous function, moreover, the discontinuity point corresponds to the complete solvation limit (CSL). The increase in the activity coefficient after CSL is determined by the dehydra-tion of the stoichiometric mixture of ions, the transition of the solvent from the inner sphere into the volume of the solution for the subsequent solvation of additional portions of the electrolyte" [31, 32]. A similar view on the dynamics of concentrated salt solutions was al-so expressed in the monograph [33] (Chapter 2.2). The presence of a non-evaporating liq-uid may be due to the fact that this liquid has reached the CSL, and there is practically no free water in it (Fig. 9).

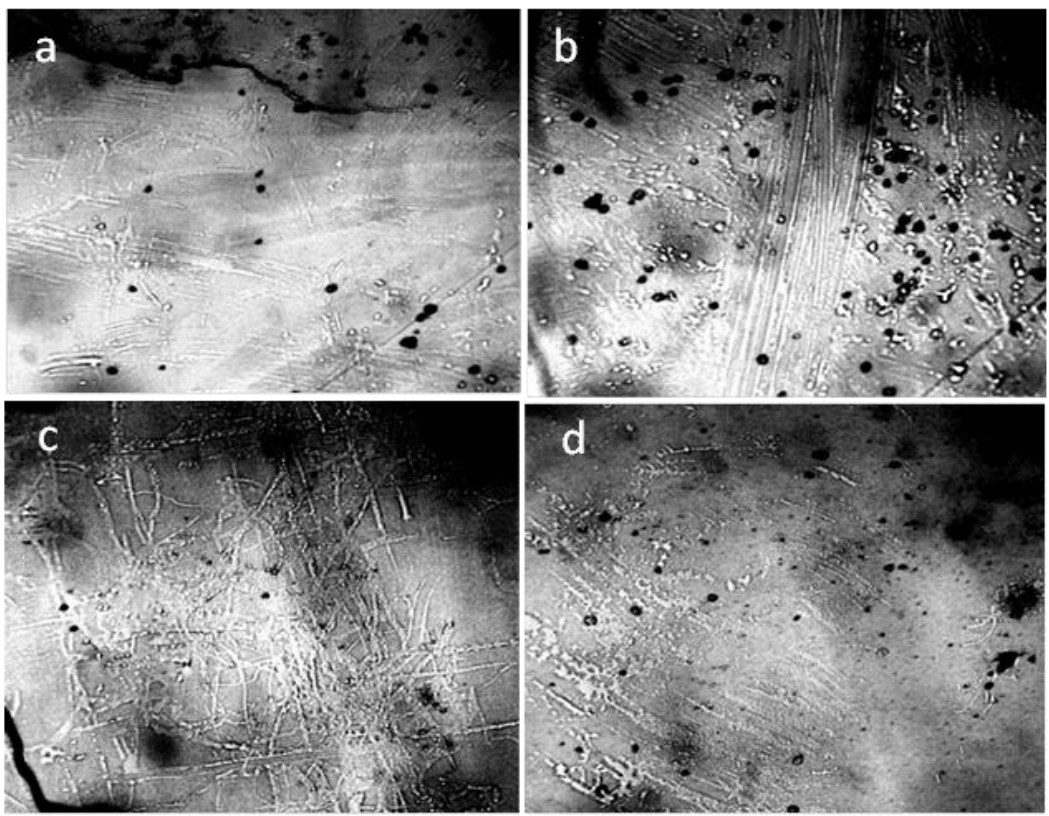

**Figure 9.** Traces of non-drying (gel-like) water on the glass surface 2 weeks after evaporation of free water: a, b – at the bottom of the Petri dish; c,d – at the bottom of slides incubated in distilled water. Dark NaCl crystals are visible. Fragment b shows a mark from the tip of a toothpick that violated the integrity of the film. Microphoto. The width of each frame is 3 mm.

No traces of non-drying water could be found in the sediment after the evaporation of tap water. The entire field of view was occupied by crystalline structures, both of fractals and in the form of monolithic NaCl crystals. (Figure 9). Two weeks later, the structure of the sediment changed (Figure 10): individual small DP aggregates became even smaller and acquired visible signs of erosion (Figure 10, a, c). Large NaCl crystals formed on a number of such aggregates. The surface of large DP aggregates (Figure 10, b, d) was covered with the same crystals. In general, it can be argued that during this time the process of DP de-struction and crystal growth continued.

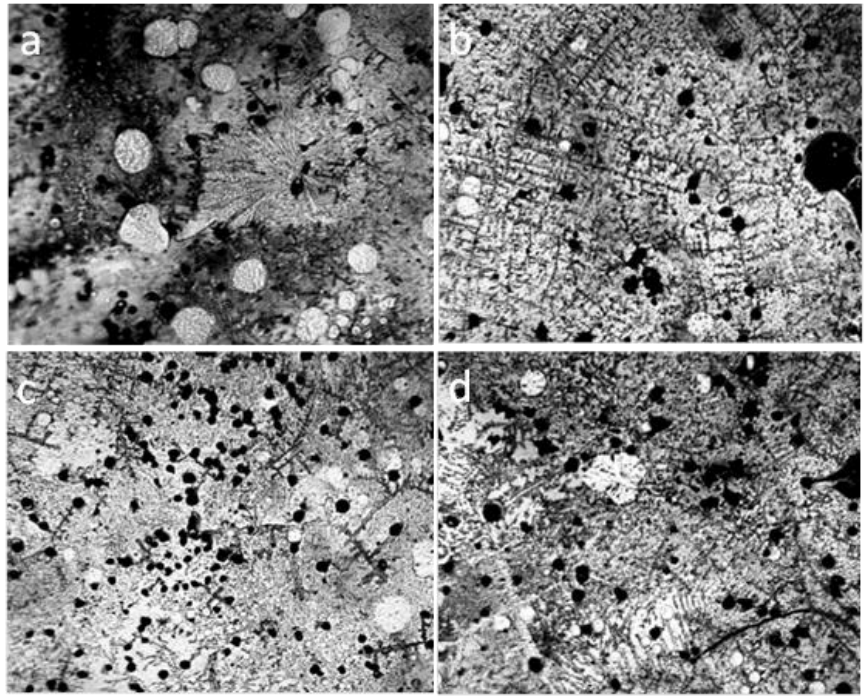

**Figure 10.** Fresh sediment of tap water in a Petri dish. One can see the remains of DP aggregates (rounded light figures - a, d), fractal salt clusters (b, c) and large NaCl crystals growing on them. Microphoto. The width of each frame is 3.0 mm.

Slow structuring of aqueous solutions may be a consequence of the "drop" structure of associated liquids, such as water, the cooperative nature of H-bonds in "drops" of the intrinsic structure of water and associates of solvates of substances dissolved in it [34].

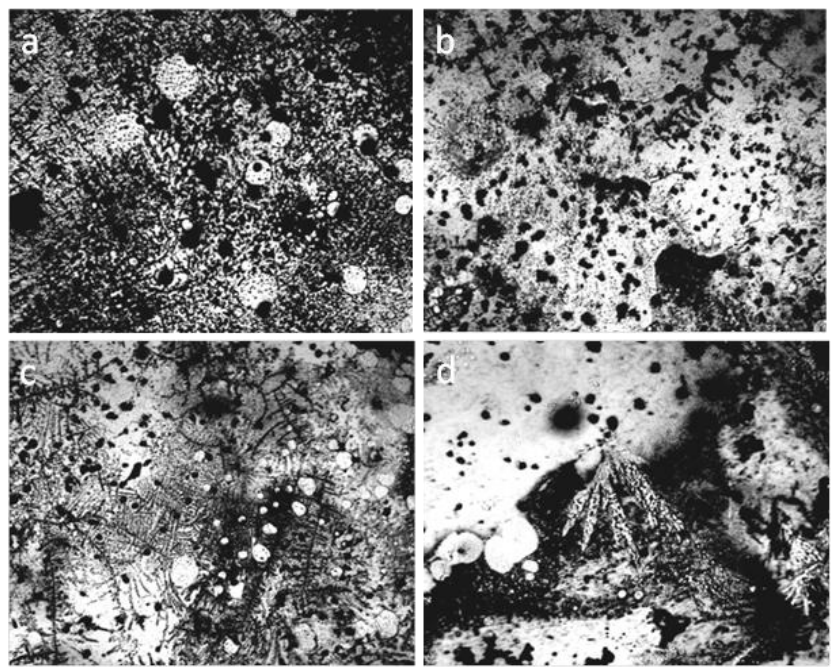

**Figure 11.** Sediment of tap water at the bottom of a Petri dish two weeks after evaporation of free water. Continued destruction of liquid crystal hydration shells in DP aggregates and growth of NaCl crystals. Microphoto. The width of each frame is 3.0 mm.

The study showed that optical microscopy is an accessible, inexpensive and informa-tive tool for studying dynamic processes in liquid media. This approach opens up a pano-ramic picture of interrelated phase transitions, which is difficult to obtain using separate "point" research methods designed to clarify the details and mechanisms of the observed phenomena. It is important that the stimulus that "starts" all this dynamics is the natural process of evaporation under room conditions.

This work confirmed our earlier observations [14, 15] and expanded them by revealing non-drying water at the end of the evaporation of distilled water. Based on the data ob-tained, we built our own picture of the observed processes. However, unresolved questions remain regarding the presence of sodium chloride in distilled water [14, 15]. We hope that this will be the subject of further research by authors interested in this topic.

**Supplementary Materials:** The following supporting information can be downloaded at:

Microstructure of tap water sediments on hydrophilic surfaces.pdf https://disk.yandex.ru/i/uQz0YUOIq5HIOg; Microstructure of distilled water sediments on hy-drophilic surfaces.pdf https://disk.yandex.ru/i/RMReWkX888aDlA

**Author Contributions:** Conceptualization, T.Y. and V.Y.; methodology, T.Y. and A.S; validation, T.Y. and A.S; formal analysis, V.Y.; investigation, T.Y., A.S.; data curation, V.Y.; writing—original draft preparation, T.Y.; writing—review and editing, T.Y, A.S., V.Y. All authors have read and agreed to the published version of the manuscript.

**Funding:** This research was funded by the Institute of Applied Physics RAS (Project No. 0030-2021-0014)

**Institutional Review Board Statement:** Not applicable.

**Informed Consent Statement:** Not applicable.

**Data Availability Statement:** Not applicable.

**Acknowledgments:** The authors are grateful to Dr. Grigory Gelikonov for stimulating this study.

**Conflicts of Interest:** The authors declare no conflict of interest.